\documentclass[12pt]{article}
\usepackage{pdproc,graphicx} 

\begin{document}

\renewcommand{\thefootnote}{\alph{footnote}}
  
\title{DISCOVERING LIGHT PSEUDOSCALAR BOSONS IN DOUBLE-PULSAR OBSERVATIONS\footnote{Talk given by M. R. at the ``Third International Workshop on Neutrino Oscillations in Venice'' (to appear in the Proceedings).} }

\author{ARNAUD DUPAYS}

\address{ Institute of Physical and Theoretical Chemistry, \\ 
Technical University of Munich,\\
D-85747 Garching, Germany\\
 {\rm E-mail: dupays@ch.tum.de}}

  \centerline{\footnotesize and}

\author{MARCO RONCADELLI}

\address{INFN, Via A. Bassi 6, I-27100 Pavia, Italy\\
{\rm E-mail: marco.roncadelli@pv.infn.it}}

\normalsize\baselineskip=15pt

\abstract{The axion is just one from a general class of new particles -- called Light Pseudoscalar Bosons (LPBs) -- predicted by many realistic extensions of the Standard Model. We offer a somewhat pedagogical review of their main properties, with particular emphasis on the effects they induce in a light beam travelling in an external magnetic field, like photon-LPB oscillations, birefringence and dichroism. Moreover, we discuss a new strategy whereby LPBs can be discovered by high-precision observations of certain binary neutron-star systems. Basically, 
in a double pulsar seen almost edge-on, photon-LPB oscillations can give rise to a characteristic attenuation pattern of the light beam emitted by one of the pulsars when it goes through the magnetosphere of the companion. Depending on the actual values of the LPB mass and its two-photon coupling constant, the effect can be seen in the $\gamma$-ray band with the upcoming GLAST mission. We also shown that this method provides a remarkable cross-check for the recent claim by the PVLAS collaboration about the existence of a new LPB.} 

\normalsize\baselineskip=15pt

\section{Introduction}

Light pseudoscalar bosons (LPBs) are predicted by many realistic extensions of the Standard Model and have attracted considerable interest in the last few years\cite{Raffelt1996}. Besides than in four-dimensional models\cite{masso1}, they naturally arise in the context of compactified Kaluza-Klein theories\cite{kk} as well as in superstring theories\cite{superstring}.

Certainly the most well-known example of LPB is the {\it axion}, namely the pseudo-Goldstone boson associated with the $U(1)$ Peccei-Quinn global symmetry invented to solve the {\it 
strong} $CP$-{\it problem}\cite{axion}.

As a rule, LPBs are very light and extremely weakly coupled to ordinary matter, so that they are quite elusive in accelerator experiments. Still, their characteristic two-photon coupling gives rise to photon-LPB conversion effects in the presence of {\it external} electromagnetic fields. This phenomenon -- quite similar in nature to flavour oscillations for massive neutrinos -- offers the unique opportunity to detect LPBs by means of high-precision non-accelerator experiments. 

In fact, the PVLAS collaboration\cite{pvlas,pvlas2} has recently reported positive evidence in favour of a LPB with mass $m \simeq 1.0 \cdot 10^{- 3} \, {\rm eV}$ and two-photon inverse coupling constant\footnote{See eq. (\ref{a5}) below.} $M \simeq 3.8 \cdot 10^{5} \, {\rm GeV}$. As we will see, this discovery would definitely signal the existence of {\it new physics} at low energy which goes far beyond LPBs, and so an independent confirmation of the PVLAS claim looks compelling. 

Besides, the two-photon coupling entails profound implications for LPBs in astrophysics and cosmology\cite{Turner}, and -- depending on their mass and interaction strength -- they can also be a natural candidate for nonbaryonic dark matter in the Universe.

Our aim is twofold. We review the present status of LPB physics, with particular attention to the phenomenology arising from their two-photon coupling. After this rather general discussion, we address a specific astrophysical implication of the photon-LPB conversion mechanism, which leads to a novel strategy whereby LPBs can be discovered by high-precision observations of certain binary neutron-star systems\cite{drrb}. Basically, in a double pulsar seen almost edge-on, photon-LPB oscillations can give rise to a characteristic attenuation pattern of the light beam emitted by one of the pulsars when it traverses the magnetosphere of the companion. Depending on the actual values of the LPB mass and its two-photon coupling constant, the effect can be seen in the $\gamma$-ray band with the upcoming GLAST mission. We also shown that this method provides a remarkable cross-check for the above PVLAS claim.

\section{QED vacuum effects}

In order to better appreciate the physical phenomena through which LPBs can be detected, it is convenient to discuss first pure QED effects on the propagation of light in a {\it magnetized} vacuum. Indeed, we shall see later that qualitatively identical effects are brought about by LPBs. Moreover, in order to disentangle the LPB signal from the observed data an estimate of the QED background is necessary.

Our starting point is the lagrangian 
\begin{equation}
\label{a1} 
{\cal L}_1 = - \frac{1}{4} \, F_{\mu \nu} \, F^{\mu \nu} + \frac{{\alpha}^2}{90 \, m^4_e} \, 
\left[ \left(F_{\mu \nu} \, F^{\mu \nu} \right)^2 + \frac{7}{4} \left(F_{\mu \nu} \, 
\tilde F^{\mu \nu} \right)^2 \right]~,  
\end{equation}
where the second term on the r.h.s. is the Heisenberg-Euler effective lagrangian\cite{Heisenberg}, which describes the one-loop corrections to classical electrodynamics -- due to the exchange of virtual fermions -- for photon frequency $\omega$ much smaller than the electron mass $m_e$. Here, $\alpha$ is the fine-structure constant, $F^{\mu \nu}$ is the usual electromagnetic field strength, $\tilde F_{\mu \nu}$ is its dual and natural Lorentz-Heaviside units with $\hbar=c=1$ are employed throughout. 

We assume that an external {\it homogeneous} magnetic field $\bf B$ is present, whose direction makes an angle $\theta$ with the momentum ${\bf k}$ of a propagating photon. We also suppose that 
$\bf B$ is much smaller than the {\it critical} magnetic field $B_{\rm cr} \equiv m_e^2/e 
\simeq 4.41 \cdot 10^{13} \, {\rm G}$.

We investigate the effects of $\bf B$ at one-loop level, to lowest nontrivial order in the coupling constant. Because of Furry's theorem, diagrams with an odd number of photon vertices vanish. So, the triangle diagram with one photon line representing the external magnetic field vanishes, and the leading contribution to the photon propagator is given by the box diagram with two external-field lines. Clearly this contribution is of order $\alpha^2 B^2$, but one factor of $\alpha$ can be absorbed into $B_{\rm cr}^2$ so that it actually goes like $\alpha \, (B/B_{\rm cr})^2$. An explicit calculation shows that the propagation eigenstates corresponding to lagrangian ${\cal L}_1$ are photons with linear polarization either parallel or normal to the plane defined by the vectors $\bf B$ and $\bf k$: they will be denoted by ${\gamma}_{\parallel}$ and ${\gamma}_{\bot}$, respectively\footnote{We follow the convention to define the photon polarization in terms of the direction of the {\it electric} field (no general consensus exists on this point).}. Owing to the box diagram under consideration, the velocity of ${\gamma}_{\parallel}$ and ${\gamma}_{\bot}$ in vacuo {\it differs} from $c$, and can be conveniently expressed in terms of the corresponding indices of refraction. Explicitly, we have\cite{Dittrich}
\begin{equation}
\label{a2}
n_{\parallel} = 1 + \frac{7}{2} \, \left( \frac{\alpha}{45 \pi} \right) \, 
\left( \frac{B \, {\rm sin} \, \theta}{B_{\rm cr}} \right)^2~,
\end{equation}
\begin{equation}
\label{a3}
n_{\bot} = 1 + \frac{4}{2} \, \left( \frac{\alpha}{45 \pi} \right) \, 
\left( \frac{B \, {\rm sin} \, \theta}{B_{\rm cr}} \right)^2~.
\end{equation}
Thus, we see that the magnetized QED vacuum produces a selective change in the velocity of light depending on its polarization, a phenomenon called {\it birefringence} in analogy with what happens in an anisotropic optical medium. Notice however that here birefringence is {\it achromatic}, since $n_{\parallel}$ and $n_{\bot}$ are independent of $\omega$.

As a matter of fact, the above box diagram is relevant also for a different process. For, suppose that we replace an external-field line with a propagating-photon line. Then what we learn is that in the magnetized QED vacuum the phenomenon of {\it photon splitting} takes place\cite{Dittrich}. Even though this argument leads to a correct conclusion, it gives a wrong feeling about the orders of magnitudes involved. Indeed, it turns out that the diagram in question vanishes in the limit of no vacuum dispersion, namely for $\omega = k$. Although the previous discussion has shown that vacuum is necessarily a dispersive medium in the presence of a magnetic field, an explicit calculation\cite{adler} reveals that the leading contribution to the photon splitting amplitude does {\it not} come from the box diagram but rather from the exagon diagram with three external-field lines, so that it goes like ${\alpha}^{3/2} \, (B/B_{\rm cr})^3$.

While the significance of birefringence is clear, further discussion is required in order to bring out the physical relevance of photon splitting. Similarly to what happens for birefringence, the amplitude for photon splitting depends on the polarization of the initial photons. A systematic analysis shows that this amplitude {\it vanishes} for ${\gamma}_{\parallel}$ photons, so that these photons do not split. Furthermore, it turns out that ${\gamma}_{\bot}$ photons split predominantly into ${\gamma}_{\parallel}$ photons\cite{adler}. Hence, we see that the magnetized QED vacuum also gives rise to a selective absorption of light depending on its polarization\footnote{We suppose that $\omega < 2 \, m_e$, which ensures that photo-pair production is kinematically forbidden (otherwise this process would dominate light absorption).}. Thanks to the analogy with what happens in an optical medium, this phenomenon is called {\it dichroism}. In particular, an unpolarized light beam becomes almost linearly polarized in the plane defined by the vectors $\bf B$ and $\bf k$, owing to photon splitting.

Actually, the foregoing considerations entail that in the QED vacuum dichroism is suppressed 
by a factor ${\alpha}^{1/2} \, (B/B_{\rm cr})$ relative to birefringence. This fact makes QED dichroism totally unobservable in terrestrial laboratories, but it becomes important in those astrophysical situations where extreme magnetic fields are known to exist, like near the surface of pulsars\cite{duncan}.

It is illuminating to illustrate the above effects in the case of a monochromatic light beam which travels in a magnetized vacuum along the $z$-direction. The beam is supposed to be {\it linearly} polarized initially (at an angle $\varphi \neq 0$ with respect to the plane defined 
by $\bf B$ and $\bf k$). 

\begin{itemize}

\item Because of birefringence, the two modes ${\gamma}_{\bot}$ and ${\gamma}_{\parallel}$ propagate with different speeds. Therefore at any finite distance from the source, the beam polarization turns out to be {\it elliptical}. Even more explicitly, as the light beam moves along the $z$-axis, its electric field changes both its direction and its magnitude so as to trace a spiral (around the $z$-axis) with elliptical section. After each $2 \pi$ rotation, a different projected ellipse is singled out in the plane perpendicular to ${\bf k}$. Still -- as long as birefringence alone is concerned -- all such ellipses have {\it parallel} major axis, which is just an elementary manifestation of the composition of two harmonic motions along orthogonal directions. 

\item Imagine now that only photon splitting is present. Then the resulting dichroism depletes the ${\gamma}_{\bot}$ mode, so that the electric field in the ${\gamma}_{\bot}$ mode gets reduced while that in the ${\gamma}_{\parallel}$ mode gets increased. Geometrically, the net result is a {\it rotation} of the electric field of the beam.

\end{itemize}
As a consequence, when a light beam produced with {\it linear} polarization travels in a magnetized vacuum an {\it elliptical} polarization shows up, with the major axis of the ellipses {\it rotated} with respect to the initial polarization.  

So far, we have supposed that the external magnetic field is homogeneous. When this is not the case, the above discussion is not altered in a qualitative way. Yet, a novel effect shows up. Indeed, a {\it varying} index of refraction can produce {\it quantum vacuum lensing}\cite{qvl}. This phenomenon is quite similar in nature to optical and gravitational lensing, apart from the fact that it depends on the light {\it polarization}. Observe however that QED quantum vacuum lensing is {\it achromatic}.

\section{Light Pseudoscalar Bosons}

As we said, LPBs are present in many realistic extensions of the Standard Model and their implications for elementary-particle phenomenology, astrophysics and cosmology have been thoroughly investigated\cite{Raffelt1996,masso1,axion,Turner}.

\subsection{General properties}

A generic feature of LPBs is a $CP$-conserving two-photon coupling, so that they are described by the lagrangian
\begin{equation}
\label{a5}
{\cal L}_2 = \frac{1}{2} \, \partial^{\mu} \phi \, \partial_{\mu} \phi - \frac{1}{2} \, m^2 \,{\phi}^2 - 
\frac{1}{4 M} \, F^{\mu \nu} \, \tilde F_{\mu \nu} \, \phi~,
\end{equation}
where $\phi$ denotes the LPB field and $M$ is the two-photon inverse coupling constant with the dimension of an energy\footnote{We stress that the following considerations apply also to a {\it scalar} light boson, provided the $F^{\mu \nu} \, \tilde F_{\mu \nu} \, \phi$ coupling is replaced by $F^{\mu \nu} \, F_{\mu \nu} \, \phi$.}.

A crucial consequence of ${\cal L}_2$ is that the propagation eigenstates of the photon-LPB system differ from the corresponding interaction eigenstates. Hence interconversion takes place, much in the same way as it happens for massive neutrinos of different flavours\cite{neutrino oscil}. However, since the mixing term in ${\cal L}_2$ involves ``two photons'', one of them must correspond to an external field. So, photon-LPB transitions occur only in the presence of {\it external} electromagnetic fields. Observe that such an external field is also needed to compensate the photon-LPB spin mismatch. 

As far as a {\it generic} LPB is concerned, its mass $m$ and two-photon inverse coupling constant $M$ are regarded as {\it independent} parameters, which are constrained by the known phenomenology (more about this, later). As an orientation, we shall assume $m < 1 \, {\rm eV}$ in the following.

The {\it axion} naturally fits within this context. Its properties are fully specified in terms of the scale $f_a$ at which the $U(1)$ Peccei-Quinn global symmetry is spontaneously broken by the vacuum expectation value of a singlet Higgs field. Explicitly, the axion mass is
\begin{equation}
\label{a6}
m \simeq 0.6 \left( \frac{10^7 \, {\rm GeV}}{f_a} \right) \, {\rm eV}~,
\end{equation}
while its two-photon inverse coupling constant is
\begin{equation}
\label{a7}
M \simeq 1.2 \cdot 10^{10} \, k \, \left( \frac{f_a}{10^7 \, {\rm GeV}} \right) \, \, {\rm GeV}~,
\end{equation}
where the constant $k$ is a model-dependent parameter roughly of order one ($k = 1$ in the DFSZ model)\cite{cgn}. Hence, the axion is characterized by the following mass-coupling relation
\begin{equation}
\label{a8}
m \simeq 0.7 \cdot k \, \left( \frac{10^{10} \, {\rm GeV}}{M} \right) \, {\rm eV}~.
\end{equation}  
Astrophysical and cosmological considerations\cite{Raffelt1996,Turner} suggest $10^{7} \, {\rm GeV} < f_a < 10^{13} \, {\rm GeV}$, which -- thanks to eqs. (\ref{a6}) and (\ref{a7}) -- entails in turn $10^{10} \, {\rm GeV} < M < 10^{16} \, {\rm GeV}$ and $10^{ - 6} \, {\rm eV} < m < 1 \, {\rm eV}$, where we have taken $k \simeq 1$.

\subsection{Photon-LPB oscillations}

At this point, it is appropriate to give a closer look at the phenomenon of photon-LPB oscillations. 

Manifestly, they are described by the {\it second-order} coupled Klein-Gordon and Maxwell equations dictated by lagrangian ${\cal L}_1 + {\cal L}_2$. Quite often -- due to the very small LPB mass -- one is interested in the regime where the photon/LPB energy $\omega$ is {\it much greater} than the LPB mass $m$ itself. In such a situation, the short-wavelength approximation can be successfully applied and turns the above wave equation into a {\it first-order} propagation equation. More specifically, we consider a monochromatic light beam travelling along the $z$-direction in the presence of an {\it arbitrary} magnetic field ${\bf B}$. Accordingly, the propagation equation takes the following form\cite{Raffelt and Stodolsky}
\begin{equation}
\label{aa6}
i \, \frac{\partial}{\partial z} | \psi (z) \rangle = {\cal M} | \psi (z) \rangle~,
\end{equation}
with
\begin{equation}
\label{aa7}
| \psi (z) \rangle \equiv c_x (z) \, |x \rangle + c_y (z) \, |y \rangle + c_{\phi} (z) \, 
| \phi \rangle~,        
\end{equation} 
where $| x \rangle$ and $| y \rangle$ are the two photon linear polarization states along the 
$x$ and $y$ axis, respectively, and $| \phi \rangle$ denotes the LPB state. In the $\{| x \rangle, | y \rangle, | \phi \rangle \}$ basis, the mixing matrix ${\cal M}$ is given by
\begin{equation}
\label{aa8}
{\cal M} = \left(
\begin{array}{ccc}
\omega + \Delta_{xx} & \Delta_{xy} & B_x /2M \\
\Delta_{yx} & \omega + \Delta_{yy} & B_y /2M \\
B_x /2M & B_y /2M & \omega - m^2 /2 \omega \\
\end{array}
\right)~.
\end{equation}
While the terms appearing in the third row and column of ${\cal M}$ have an evident physical meaning, the $\Delta$-terms require some explanation. Generally speaking, they reflect both the properties of the medium and the QED vacuum effects addressed in Section 2. Off-diagonal 
$\Delta$-terms directly mix the photon polarization states and typically give rise to Faraday rotation. 

When the external magnetic field is {\it homogeneous}, we can choose the $y$-axis along the 
projection of ${\bf B}$ perpendicular to the $z$-axis. Correspondingly we have $B_x = 0$, $B_y = B \, {\rm sin} \, {\theta}$, $A_x = A_{\bot}$, $A_y = A_{\parallel}$, and the mixing matrix takes the form
\begin{equation}
\label{a9}
{\cal M} = \left(
\begin{array}{ccc}
\omega + \Delta_{\bot} & \Delta_R & 0 \\
\Delta_R & \omega + \Delta_{\parallel} & B \, {\rm sin} \, {\theta} /2M \\
0 & B \, {\rm sin} \, {\theta} /2M & \omega - m^2 /2 \omega \\
\end{array}
\right)~.
\end{equation}
In general, the diagonal $\Delta$-terms receive three different contributions, and so we write
\begin{equation}
\label{a10}
{\Delta}_{\parallel, \bot} = {\Delta}_{\parallel, \bot}^{\rm QED} + {\Delta}_{\parallel, 
\bot}^{\rm PL} + {\Delta}_{\parallel, \bot}^{\rm CM}~.
\end{equation}
The terms ${\Delta}_{\parallel, \bot}^{\rm QED}$ represent the QED vacuum effects. They follow directly from eqs. (\ref{a2}), (\ref{a3}) and read
\begin{equation}
\label{a11}
{\Delta}_{\parallel}^{\rm QED} = \left( n_{\parallel} - 1 \right) \omega = \frac{7}{2} \, \left( \frac{\alpha}{45 \pi} \right) \, \left( \frac{B \, {\rm sin} \, \theta}{B_{\rm cr}} 
\right)^2 \omega~,
\end{equation}
\begin{equation}
\label{a12}
{\Delta}_{\bot}^{\rm QED} = \left( n_{\bot} - 1 \right) \omega = \frac{4}{2} \, \left( \frac{\alpha}{45 \pi} \right) \, \left( \frac{B \, {\rm sin} \, \theta}{B_{\rm cr}} 
\right)^2 \omega~. 
\end{equation}
In the presence of a plasma, charge screening effects produce an effective photon mass given by the plasma frequency $\omega_{\rm pl}^2 \equiv 4 \pi \alpha \, n_e/ m_e$ with $n_e$ denoting the electron density, and the resulting contribution is 
\begin{equation}
\label{a13}
{\Delta}^{\rm PL}_{\parallel, \bot} = - \, \frac{\omega_{\rm pl}^2}{2 \omega} = - \, 
\frac{2 \pi \alpha \, n_e}{m_e \, \omega}~.
\end{equation}
Furthermore, the term ${\Delta}_{\parallel, \bot}^{\rm CM}$ describes the Cotton-Mouton effect, namely the birefringence of a fluid due to a transverse magnetic field. Finally, the term 
$\Delta_R$ accounts for Faraday rotation. 

In the situations to be discussed below, both ${\Delta}_{\parallel, \bot}^{\rm CM}$ and 
$\Delta_R$ turn out to be irrelevant, and so they will be discarded from now on. Consequently, the longitudinal component of the magnetic field $B \, {\rm cos} \, {\theta}$ disappears from ${\cal M}$, $A_{\bot}$ decouples while only $A_{\parallel}$ mixes with $\phi$. As a result, the mixing matrix ${\cal M}$ reduces to the two-dimensional form
\begin{equation}
\label{a14}
{\cal M}_0 = \left(
\begin{array}{cc}
\omega + \Delta_{\parallel} & B \, {\rm sin} \, {\theta} /2M \\
B \, {\rm sin} \, {\theta} /2M & \omega - m^2 /2 \omega \\
\end{array}
\right)~.
\end{equation}
As is well known, such a matrix can be diagonalized by an orthogonal transformation with rotation angle
\begin{equation}
\label{a15}
\Theta = \frac{1}{2} \, {\rm arctg} \left( \frac{B \, {\rm sin} \, \theta /M}{\Delta_{\parallel} + m^2/2 \omega} \right)
\end{equation}
and -- in complete analogy with the case of neutrino oscillations\cite{neutrino oscil} -- the probability the a ${\gamma}_{\parallel}$ photon will be converted into a LPB after travelling a distance $d$ is 
\begin{equation}
\label{a16}
P ({\gamma}_{\parallel} \to {\phi}) = {\rm sin}^2 2 \Theta \  {\rm sin}^2 
\left( \frac{\Delta_{\rm osc} \, d}{2} \right)~,
\end{equation}
where the oscillation wavenumber is
\begin{equation}
\label{a17}
{\Delta}^2_{\rm osc} = \left( \Delta_{\parallel} + \frac{m^2}{2 \omega} \right)^2 + 
\left( \frac{B \, {\rm sin} \, \theta}{M} \right)^2~,
\end{equation}
so that the  oscillation length is just $L_{\rm osc} = 2 \pi / {\Delta}_{\rm osc}$. Notice that 
$P ({\gamma}_{\parallel} \to {\phi})$ becomes energy-independent when ${\Delta}_{\rm osc}$ is dominated by the photon-LPB mixing term.

\subsection{Photon propagation}

A further consequence of the photon-LPB mixing concerns the propagation of light. As first pointed out by Maiani, Petronzio and Zavattini\cite{Maiani}, this actually comes about in two distinct ways. 

We have seen in Section 2 that the exchange of virtual fermions yields a nontrivial contribution to the photon propagator in the presence of an external magnetic field, which affects the velocity of light and gives rise to vacuum birefringence. A further contribution of the same kind arises from the exchange of virtual LPBs, namely from the diagram in which two $\gamma \gamma \phi$ vertices are joined by the $\phi$ line and two $\gamma$ lines represent the external field. Again, the corresponding amplitude depends on the polarization of the initial photon, and so it is an {\it additional} source of vacuum {\it birefringence}. As before, the photon propagation eigenstates are the linear polarization modes ${\gamma}_{\parallel}$ and ${\gamma}_{\bot}$, but a small complication arises here, due to the fact that the mass eigenstates differ from the interaction eigenstates. More specifically, ${\gamma}_{\parallel}$ and ${\gamma}_{\bot}$ now denote the polarization modes of the photon states which diagonalize the mixing matrix ${\cal M}$. We suppose for simplicity that the magnetic field is {\it homogeneous}. Then we know that the ${\gamma}_{\bot}$ mode decouples from $\phi$, and hence its refractive index is still given by eq. (\ref{a3}). On the other hand, a straightforward calculation\cite{Raffelt and Stodolsky} shows that eq. (\ref{a2}) gets presently replaced by
\begin{equation}
\label{a18}
\vbox{\displaylines{n^\prime_{\parallel} = 1 + \frac{7}{2} \, \left( \frac{\alpha}{45 \pi} \right) \, 
\left( \frac{B \, {\rm sin} \, \theta}{B_{\rm cr}} \right)^2 \cr 
+ \frac{1}{2 \omega} \left\{ 
\left[ \left( \frac{B \, {\rm sin} \, \theta}{M} \right)^2 + \left( {\Delta}_{\parallel} + 
\frac{m^2}{2 \omega} \right)^2 \right]^{1/2} - \left( {\Delta}_{\parallel} + 
\frac{m^2}{2 \omega}   \right) \right\}~.\cr}} 
\end{equation}
Since we expect a similar result to hold true also when the magnetic field is inhomogeneous, we see that the resulting quantum vacuum lensing becomes {\it chromatic} in the presence of LPBs. 

Next, we address the effect arising from the production of real LPBs, occurring via the $\gamma \gamma \phi$ vertex with a photon line representing the external field. Because only ${\gamma}_{\parallel}$ photons mix with LPBs, photon-LPB conversion is an {\it additional} source of vacuum {\it dichroism}. 

Consider now a monochromatic light beam which travels in a magnetized vacuum along the $z$-direction, and suppose that it is {\it linearly} polarized initially (at an angle $\varphi \neq 
0$ with respect to the plane defined by $\bf B$ and $\bf k$). How is its propagation affected by the existence of LPBs? We know that birefringence gives rise to an {\it elliptical} polarization, whereas dichroism produces a {\it rotation} of the ellipse's major with respect to the initial polarization. Manifestly, these effects are qualitatively identical to those arising from the QED magnetized vacuum, but the point to be stressed here is that a measurement of both the beam ellipticity and the rotation angle -- induced by LPBs -- permits a {\it complete} determination of their two independent parameters, namely $m$ and $M$. Such a circumstance turns out to be of immediate relevance for the discovery of LPBs, as we will see in Section 4.

\subsection{Astrophysical constraints}

A crucial implication of the two-photon coupling of LPBs concerns stellar evolution. Thermal photons in the inner regions of stars are easily converted into LPBs in the fluctuating electromagnetic fields of the stellar plasma. So long as LPBs are fully described by lagrangian 
${\cal L}_2$, they traverse unimpeded the stellar material and escape into free space. Hence, the stellar energy-loss rate gets increased in the presence of LPBs. Because of the negative specific heat resulting from virial equilibrium, the central temperature also increases, thereby changing the observed properties of stars. Yet, current models of stellar evolution are in fairly good agreement with observations. Therefore, the two-photon inverse coupling constant $M$ has to be large enough to provide a sufficient suppression of unwanted LPB effects.  

This argument has been applied in a quantitative fashion to stars of different kinds\footnote{We do not need to be more precise about this point here.}, with the result\cite{Raffelt1996} that typically $M > 10^{10} \, {\rm GeV}$.

Remarkably enough, basically the same conclusion is reached by the CAST experiment, designed to detect LPBs (axions) emitted by the Sun~\cite{cast} (more about this, later).

\section{Laboratory experiments}

Over the last two decades, various proposals have been put forward to detect axions through their two-photon coupling in non-accelerator situations. Generic LPBs can also be discovered by these techniques, provided of course that their mass and inverse two-photon coupling constant fall into suitable ranges dictated by the experimental setup. 

A strategy addresses the possibility of converting actually existing axions into photons. As suggested by Sikivie\cite{Sikivie}, Galactic dark matter axions are expected to excite a proper TM mode of a {\it tunable} microway cavity -- permeated by a strong magnetic field -- when resonance takes place, namely when the characteristic frequency of a cavity TM proper mode 
happens to coincide with the axion mass. So far, no positive signal has been reported by experiments of this kind\cite{Bradley}. Although this method can also be applied to axions coming from the Sun, they can be more easily detected by back conversion into $X$-ray photons in a ``magnetic telescope''\cite{vanbibber}. Recently, the latter technique has been implemented by the above-mentioned CAST experiment\cite{cast}, leading to the lower bound $M > 0.86 \cdot 10^{10} \, {\rm GeV}$ for $m < 0.02 \, {\rm eV}$.

An alternative strategy relies upon virtual LPB effects produced on photon propagation. As we have seen, an initially {\it linearly}-polarized light beam travelling in a strong magnetic field is expected to become {\it elliptically} polarized, with the ellipse's major axis {\it rotated} with respect to the initial polarization. Although an induced ellipticity is also due to the magnetized QED vacuum\footnote{No observable rotation is expected to arise in the QED vacuum because dichroism is suppressed with respect to birefringence, as pointed out in Section 2.}, Maiani, Petronzio and Zavattini have shown that the LPB contribution dominates for the physically interesting region of the $m-M$ parameter plane in the case of the axion\cite{Maiani}. Recalling the conclusion reached at the end of Subsection 3.3, an independent determination of $m$ and $M$ can be achieved by measuring the axion-induced beam ellipticity and rotation angle. Needless to say, also LPBs can be searched for in this way. At variance with the previous technique, here it is clearly irrelevant whether LPBs are dark matter particles or whether they are emitted by stars. 

Using an experimental setup based on the latter method, a few years ago the lower bound $M > 2.8 \cdot 10^{6} \, {\rm GeV}$ was established for $m < 10^{ - 3} \, {\rm eV}$\cite{cameron}. Quite recently -- by exploiting a similar technique -- the PVLAS collaboration has reported positive evidence for an anomalously large value of the rotation angle in an initially linearly-polarized laser beam undergoing multiple reflection in a $5 \, T$  magnetic field\cite{pvlas,pvlas2}. In addition, also the beam ellipticity has been determined. Assuming that the effect is indeed brought about by a LPB, the corresponding physical parameters turn out to be in the range $ 1.0 \cdot 10^{- 3} \, {\rm eV} \leq m \leq 1.5 \cdot 10^{- 3} \, {\rm eV} $ and $2 \cdot 10^{5} \, {\rm GeV} \leq M\leq 6 \cdot 10^{5} \, {\rm GeV}$. In the following, we will adopt the 
values\cite{pvlas2} $m \simeq 1.0 \cdot 10^{- 3} \, {\rm eV}$ and $M \simeq 3.8 \cdot 10^{5} \, {\rm GeV}$. 

Manifestly, a look back at eq. (\ref{a8}) shows that the LPB in question {\it cannot} be the axion. Moreover, the quoted value of $M$ {\it violates} both the astrophysical bound discussed in Subsection 3.4 and the CAST result by about five orders of magnitudes. The only way out of this difficulty is evidently to suppose that such LPBs are {\it not} emitted by stars, which amounts to say that either they are not produced at all or else they are confined inside the star interiors in a manner consistent with the observed properties. In either case, {\it new physics} at energy as low as a few ${\rm KeV}$ is required. This is an exciting possibility which is becoming to be addressed\cite{masso2005}. At any rate, the need for {\it independent} tests of the PVLAS result looks imperative. 

Concerning the latter point, a few options have been contemplated. In the first place, an experiment similar to PVLAS can be performed, however with a different magnet and a laser beam with a different frequency, like in the BMV project\cite{askenazy}. A somewhat different method exploits the idea of {\it photon regeneration}\footnote{Suppose that a light beam is shone across a magnetic field, so that some photons are converted into LPB. If a screen is put on the beam path, photons are completely absorbed, but LPBs are not. So, they emerge on the other side of the screen and undergo photon back conversion if another magnetic field is present. Therefore, the presence of photons on the latter side of the screen would be an unambiguous signal of 
photon-LPB conversion\cite{Sikivie,photreg}.} in connection with a synchrotron $X$-ray beam produced in a free-electron laser, like the LCLS at SLAC and the XFEL at DESY\cite{rrs}. Alternatively, the value of $M$ found by PVLAS can be tested by data which will become available in the near future at $e^+ e^-$ colliders\cite{kleban}, since in these machines the two-photon coupling of a LPB gives rise to events with a single photon and missing energy in the final state.

\section{Double-pulsar observations}

A remarkable astrophysical implication of the photon-LPB conversion mechanism concerns a binary neutron-star system\cite{drrb} and can be illustrated as follows.

Consider a binary pulsar seen almost edge on and focus the attention on the light beam coming from one of the two stars, to be referred to as ``pulsar $A$''. Suppose further that both components are neutron stars, so that basically no accretion phenomenon perturbs the system. Denote by $\rho$ the impact parameter of the beam, namely its projected distance from the companion, that is to say from ``pulsar $B$''. When $\rho$ is large, nothing interesting is expected to happen. But when $\rho$ is small, the beam goes through the magnetosphere of pulsar 
$B$, so that the observed photons traverse a region where a strong magnetic field is present. As a consequence, some of these photons are expected to be converted into LPBs, thereby producing a characteristic attenuation pattern of the beam intensity.

Realistically, we consider the recently-discovered binary pulsar system J0737-3039\cite{DoublePulsar}, which is particularly well suited for our purposes. Indeed, its orbital plane 
has currently an inclination angle $i \simeq 87^0$, and the orbital precession of the system allows us to predict that the value $i \simeq 90^0$ could be attained before 2020. The orbital period is 2 hours and 45 minutes, while the spinning period of pulsar $B$ is 2.77 seconds.  Although the current minimum value of the impact parameter of the light beam from pulsar $A$ is somewhat uncertain, an estimate\cite{coles} yields $\rho \simeq 4 \cdot 10^3 \, {\rm km}$ and this is the value used in our analysis. Depending on the actual values of $m$ and $M$, the beam attenuation can be substantial. We stress that the effect in question does not depend on any new physics responsible for the confinement of the LPBs in the star interiors.

Rather than carrying out a general analysis for arbitrary values of $m$ and $M$, we prefer to focus here the attention on the specific values corresponding to the PVLAS result, namely 
$m \simeq 1.0 \cdot 10^{- 3} \, {\rm eV}$ and $M \simeq 3.8 \cdot 10^{5} \, {\rm GeV}$. 

The geometry of the considered double pulsar is depicted in Fig. 1. We introduce a $(x,y,z)$ coordinate system fixed with pulsar $B$. The light beam from pulsar $A$ propagates along the $z$ direction and passes pulsar $B$ at impact parameter $\rho$. For simplicity, we suppose that the rotation axis of pulsar $B$ lies along the $y$ direction and that its magnetic dipole moment precedes at $45^0$ about the $y$ axis. Therefore, the light beam from pulsar $A$ experiences a time-dependent and inhomogeneous dipolar magnetic field ${\bf B}$, with typical strength $B_0 \simeq 10^{12} \, {\rm  G}$ on the surface of pulsar B. In such a situation, the mixing matrix has the general form (\ref{aa8}). Since we are not interested in the polarization of the photon beam, we can safely set 
\begin{equation}
\label{b1}
\Delta_{xy} \simeq  \Delta_{yx} \simeq 0~. 
\end{equation}
For the same reason, we can take 
\begin{equation}
\label{b2}
\Delta_{xx}^{\rm QED}~ \simeq  \Delta_{yy}^{\rm QED} \simeq \left( \frac{\alpha}{45 \pi} \right) \, \left( \frac{B}{B_{\rm cr}} \right)^2 \omega~, 
\end{equation}
on account of eqs. (\ref{a11}) and (\ref{a12}). Plasma effects arise from the presence of free charges in the magnetosphere of pulsar $B$ and their contribution is given by eq. (\ref{a13}) with the electron density supplied by the Goldreich-Julian relation\cite{goldreich julian}
\begin{equation}
\label{b3}
n_e \simeq 7 \cdot 10^{- 2} \left( \frac{B}{G} \right) \left(  \frac{s}{P} \right) \, {\rm cm^{- 3}}~, 
\end{equation}
where $P$ denotes the spinning period of pulsar $B$. 

\begin{figure}[h]
\begin{center}
\includegraphics[width=8cm]{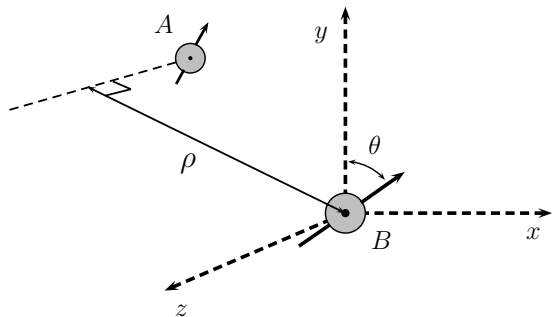}
\caption{\label{fig0}{Geometry of the model binary pulsar system}}
\end{center}
\end{figure}

We proceed to solve the resulting eq. (\ref{aa6}) numerically in the equivalent exponential form 
\begin{equation}  
\label{b4}  
| \psi(z) \rangle = {\cal T} \exp \left( -i \int_{z_0}^{z} dz^\prime {\cal M}(z^\prime) \right) 
| \psi(z_0) \rangle~,  
\end{equation} 
where ${\cal T}$ is Dyson's ordering operator. Since the ${\cal M}$ matrix is explicitly 
$z$-dependent, the propagation is computed iteratively, i.e. in steps of small intervals $\delta
z$. For small enough $\delta z$, the ${\cal M}$ matrix can be considered as constant over a single distance step, and the short-propagation iterations can be calculated as
\begin{equation}
\label{b5}
| \psi (z + \delta z) \rangle \simeq \exp \Bigl( - \, i \, {\cal M}(z) \, \delta z \Bigr) 
| \psi(z) \rangle~.
\end{equation}
The propagation is performed from $z_0$ to $z_{\rm max}$, with the initial condition corresponding to a pure photon state. Here, $z_0$ is fixed by the trajectory of pulsar $A$ and 
$z_{\rm max}$ is chosen in such a way that $B_x(z) /2M \simeq B_y(z) /2M \simeq 0$ for 
$z>z_{\rm max}$. The transition probability is then 
\begin{equation}
\label{b6}
P({\gamma}_j \to \phi) = | \langle \phi | \psi(z_{\rm max}) \rangle |^2 = 
|c_\phi(z_{\rm max})|^2~,
\end{equation}
where $j=x$ ($y$) refers to the initial photon polarization along the $x$-axis
($y$-axis). Clearly, the transition probability for an {\it unpolarized} photon beam is  
\begin{equation}
\label{b7}
P(\gamma \to \phi) = \frac{P({\gamma}_x \to \phi)}{2}+\frac{P({\gamma}_y \to \phi)}{2}~.
\end{equation}

We plot $P(\gamma \to \phi)$ versus photon energy $\omega$ in Fig. 2, for the beam impact parameter $\rho = 4 \cdot 10^3 \, {\rm km}$.

\begin{figure}[h]
\begin{center}
\includegraphics[width=8cm]{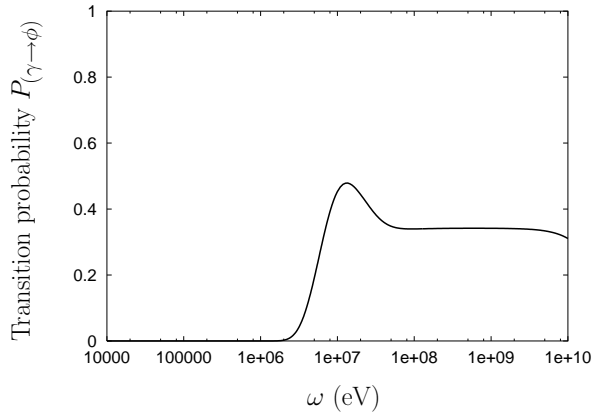}
\caption{\label{fig1}{Transition probability versus photon energy for a trajectory of the light beam with an impact parameter $\rho=4 \cdot 10^3$ km.}}
\end{center}
\end{figure}

We see that photon-LPB conversion turns out to be important for $\omega > 10 \, {\rm MeV}$, namely in the $\gamma$-ray band. This is a remarkable conclusion, since pulsar $A$ in J0737-3039 is expected to be a $\gamma$-ray source. In addition, $\gamma$-ray photons propagate totally unimpeded in the magnetosphere of pulsar $B$, and so we do not have to bother about further potential sources of beam attenuation\footnote{We stress that in the present situation photo-pair production is totally negligible.}. Finally, the short-wavelength approximation -- upon which eq. (\ref{aa6}) is based -- now appears to be fully justified. 

Actually, the behaviour of $P(\gamma \to \phi)$ in Fig. 2 can be understood in an intuitive fashion. For, imagine a drastically simplified situation in which the magnetic field experienced by the light beam is stationary and homogeneous. As already pointed out, a typical value of the magnetic field strength on the surface of a neutron star -- namely at a distance $r \simeq 10 \, {\rm km}$ from the centre -- is $B_0 \simeq 10^{12} \, {\rm  G}$. Since ${\bf B}$ is dipolar, it goes like $\sim r^{- 3}$, and so the magnetic field strength at $\rho \simeq 4 \cdot 10^{3} \, {\rm km}$ is\footnote{With a magnetic field of this strength, photon splitting is completely irrelevant.} $B \simeq 1.6 \cdot 10^{4} \, {\rm G}$. Correspondingly, from eq. (\ref{b2}) we have ${\Delta}^{QED}_{xx} \simeq {\Delta}^{QED}_{yy} \simeq 6.7 \cdot 10^{ - 18} \, (\omega /{\rm MeV}) \, {\rm eV}$.  Owing to eqs. (\ref{a13}) and (\ref{b3}), it is straightforward to see that the term ${\Delta}^{PL}_{xx} \simeq {\Delta}^{PL}_{yy}$ is much smaller than ${\Delta}^{QED}_{xx} \simeq {\Delta}^{QED}_{yy}$ for $B \simeq 1.6 \cdot 10^{4} \, {\rm G}$, and so it can be neglected. Moreover, we find $B_x / 2 M \simeq B_y/2 M \simeq 0.4 \cdot 10^{- 12} \, {\rm eV}$ and $m^2 / 2 \omega \simeq 0.5 \cdot 10^{- 12} \, ({\rm MeV}/ \omega) \, {\rm eV}$. Because the terms $\Delta_{xy}, \Delta_{yx}$ presently get only a QED contribution, they should not exceed ${\Delta}^{QED}_{xx} \simeq {\Delta}^{QED}_{yy}$. As a consequence, condition (\ref{b1}) appears to be justified. Now, a look back at eqs. (\ref{a15}) and (\ref{a16}) shows that mixing effects in ${\cal M}_0$ should become physically important for $\Theta \sim 1$. Therefore, substantial photon-LPB conversion is expected to take place for $\omega > 1 \, {\rm MeV}$, which explains the threshold behaviour of $P(\gamma \to \phi)$. Furthermore, we know that this quantity becomes energy-independent when the photon-LPB mixing term is much larger than the others. This situation presently occurs for $\omega \gg 1 \, {\rm MeV}$, thereby explaining also the flat behaviour of $P ({\gamma} \to {\phi})$.

Assuming that the light beam emitted by pulsar $A$ is not polarized, the temporal behaviour of the considered effect is best expressed in terms of the total {\it transmission} $A = 1 - P(\gamma \to \phi)$ of the beam after propagation in the magnetosphere of pulsar $B$. We plot $A$ versus time in Fig. 3, as pulsar $A$ moves in its nearly edge-on orbit. 

\begin{figure}[h]
\begin{center}
\includegraphics[width=8cm]{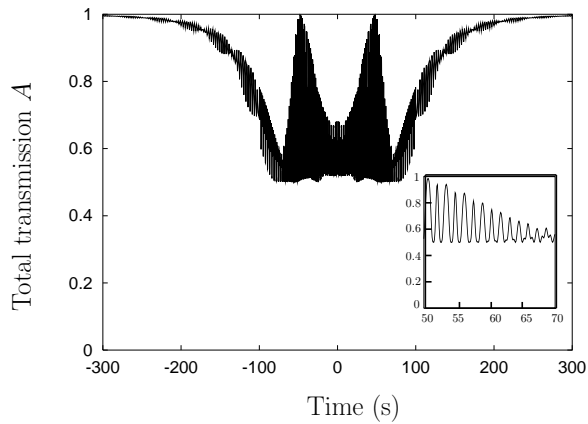}
\caption{\label{fig2}{Total transmisson of the gamma photon beam emitted by
    pulsar $A$ versus
    time. Inset shows the modulation mainly due to the rotation of the
    magnetic dipole moment of pulsar $B$.}}
\end{center}
\end{figure}

Our numerical simulation predicts a strong attenuation of the photon beam up to $50\,\%$ with a time duration of about $200$ s. As it is clear from Fig. 3, this effect has three different temporal structures. The broad minimum -- from $-200$ s to $+200$ s -- evidently corresponds to the transit of pulsar $A$ behind pulsar $B$. The tens-of-seconds, symmetric peaks are due to photon-LPB oscillations, depending on the actual path through the interaction region with pulsar $B$. Finally, the highest frequency modulation -- shown in inset -- is due to the rotation of the magnetic dipole moment of pulsar $B$. 

We note that pulsar $A$ also generates a similar effect on the light beam emitted by pulsar $B$. Since the spin period of pulsar $A$ is smaller than the one of pulsar $B$ ($23$ ms), the resulting modulation period will be also smaller.

Let us finally demonstrate that the effect under consideration is {\it observable} with the upcoming GLAST mission, which will have a pulsar $\gamma$-ray detection sensitivity up to 2 orders of magnitude better than any previous telescope\cite{Thompson}. To this end, we show in Fig. 4 the region of the $m-M$ parameter plane excluded by the no detection of an attenuation $A$ at the $10 \, \%$ level. 

\begin{figure}[h]
\begin{center}
\includegraphics[width=8cm]{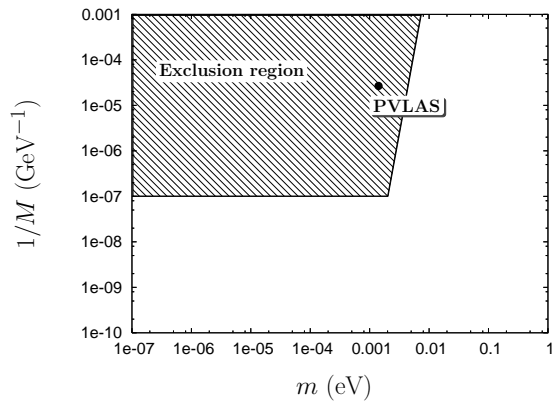}
\caption{\label{fig3}{Exclusion region in the case that the existence of the attenuation is excluded at $10\,\%$ level.}}
\end{center}
\end{figure}
Such an attenuation is achieved by a 100 photon count during the total integration time. For a two weeks observation time, this corresponds to a flux from pulsar $A$ of about $2 \cdot 10^{-7}$ photons/cm$^2$/s, which is a reasonable flux according to previous observations of several pulsars\cite{Bignami:96}. As a matter of fact, the GLAST sensitivity curves allow for a much weaker minimum detectable flux\cite{Glastlat}, down to $1 \cdot 10^{-10}$ photons/cm$^2$/s. 

Thus, we conclude that the photon-LPB conversion mechanism in the double pulsar J0737-3039 
really provides a cross-check for the recent PVLAS claim about the existence of a new 
LPB. We stress that our result would remain practically unchanged even if we were considering a light {\it scalar} boson\cite{drrb}.

\section{Acknowledgments}

We thank our collaborators Nanni Bignami and Carlo Rizzo, with whom the work reported in Section 5 was done. One of us (M. R.) would also like to thank professor Milla Baldo Ceolin for her kind invitation to talk at this splendid workshop.

\end{document}